\documentstyle[graphicx,12pt]{elsart} 

\begin{document}

\date{\today}

\begin{frontmatter}

\title{Neck fragmentation reaction mechanism}

\author[catania,bucharest]{V. Baran}
\author[catania]{M. Colonna}
\author[catania]{M. Di Toro}

\address[catania]{Laboratori Nazionali del Sud INFN, I-95123 Catania, Italy\\
 Physics \& Astronomy Dept., Univ. of Catania }

\address[bucharest]{NIPNE-HH and Physics Faculty, University of Bucharest, Romania\\
}



\begin{abstract}
Based on a microscopic transport model, we study the origin of 
nonstatistical Intermediate Mass Fragment ($IMF$) production in semicentral
heavy ion collisions at the Fermi energies. We show that a fast, dynamical
$IMF$ formation process, the {\it neck fragmentation mechanism},
can explain the experimentally observed features:
deviations from Viola systematics and anisotropic, 
narrow angular distributions. 
It may be regarded as the continuation of the multifragmentation mechanism
towards intermediate impact parameters. Its relation to other 
dynamical mechanisms, the induced
fission and the abrasion of the spectator zones, that can also contribute to
mid-rapidity $IMF$ production, is discussed. The dependence on
beam energy and centrality of the collision is carefully analysed.
The competition between volume and surface instabilities makes
this mechanism very sensitive to the in-medium nucleon-nucleon
interactions, from the cross sections for hard collisions to the
compressibility and other Equation of State ($EOS$) properties.
  
For charge asymmetric collisions the sensitivity of various observables 
to the symmetry energy is investigated. Of particular interest appears the
isospin diffusion dynamics with no signal of isospin equilibration.
However, in spite of the short time scales and of the 
dynamical aspects, we notice isoscaling features of the neck 
mechanism. We observe that isospin enrichement of the neck zone as well as 
the isoscaling parameters are sensitive to
the density dependence of asymmetry term of EOS around and below
saturation value.
\end{abstract}

\maketitle

\end{frontmatter}

\section{INTRODUCTION}

The Fermi energy domain is the transition
region between a dynamics driven by the mean-field,  below $15-20~AMeV$, 
and one where the nucleon-nucleon collisions play a central role, 
above $100~AMeV$.
It is the place of the rise and/or fall of new reaction mechanisms, including 
a liquid-gas like phase transition. Consequently it still attracts 
considerable theoretical and experimental efforts.
One of its distinctive particularities is the enhanced production of
Intermediate Mass Fragments $(IMF, 3<Z<21)$. Their characteristics
are essential to assert which dissipation mechanisms act at these
energies as well as in establishing the equilibration time hierarchy 
of various degrees of freedom. 

In particular, along the last decade, several experiments 
have been devoted to a better understanding of the origin 
of $IMF$'s  in semi-central heavy-ion collisions at bombarding energies
between $20$ and $80-100~AMeV$. From peripheral to semi-central collisions it
has been established that the reaction mechanism has a mainly binary character
and the statistical decay products from highly excited projectile-like 
(${PLF}^*$)
and targetlike (${TLF}^*$) are the first to be considered
in a consistent description. However, experimental evidences for 
$IMF$' production not directly related to the statistical decay of PLF
or TLF  were accumulated in the
past \cite{mon94}, \cite{tok95}, \cite{luk97}, \cite{sob97}
 and with the advent of $4 \pi$
arrays generations a  more systematic analysis has been possible
\cite{boc00}, \cite{lef00}, \cite{mil01}, \cite{mil02},
\cite{dav02}, \cite{pia02}, \cite{col03}.
Consequently some definite particularities of this ``dynamical'' 
$IMF$ production
have been  established:  

- a clear enhanced emission is localized in the mid-rapidity region,
intermediate between $PLF$ and $TLF$ rapidities,
especially for fragments with charge $Z$ from $3$ 
to $12$;

- the $IMF$'s relative velocity distributions with respect to $PLF$ 
(or $TLF$)
cannot be explained in terms of pure Coulomb repulsion following a 
statistical decay. A high degree of decoupling from the $PLF$ ($TLF$)
is also invoked.

- clear anisotropic $IMF$'s angular distributions are indicating preferential
emission directions and an alignment tendency;

- for charge asymmetric systems the light particles and $IMF$ emissions
 keep track of a neutron enrichment 
process that takes place in the neck region.

However, a fully consistent physical picture of the processes 
that can reproduce observed characteristics is still
a matter of debate and several physical phenomena are
taken into account.

 One is the formation of a transient, necklike structure,
joining projectile-like ($PLF$) and target-like ($TLF$).
Its noncompactness, a large surface to volume ratio, was
supposed to favor fragment emission. The creation of a cylindrical geometry
can trigger the Rayleigh instabilities which, as it happens in the
low energy fission, will drive a multiple 
neck rupture. The early decoupling of this structure may explain
qualitatively the observed trends.

A fission like mechanism can also
be included in the discussion. Indeed, after the interaction stage,
an outgoing primary nucleus,$PLF$- and/or $TLF$-like, strongly deformed 
and highly excited 
can follow a fast fission, dynamically induced, path.
A prefragment with a given energy above the barrier may form on its
deformed side. This will explain the observed deviations 
from Viola systematics. An aligned emission, due to a shorter
life before scission, is also predicted.

It is possible that with increasing bombarding energy an
abrasion-ablation  process manifests as a precursor to the
participant-spectator scenario present at higher energies.
This process will induce a third decaying hot zone that amplifies
the mid-rapidity fragment production. Related to this,
we may refer also to the recently extended  Goldhaber 
clustering model, allowing mixing of projectile
and target nucleons and including hard-scattered nucleons from
Pauli-allowed collisions, \cite{luk03}.

Even statistical decay of a hot source at intermediate energy, 
triggered by the proximity configuration with PLF and TLF,
was claimed \cite{bot99}. Finally we have to remind that  
dynamical transport models  
suggest since long time the possibility of such phenomena \cite{col95},  
\cite{dem96}, \cite{luk97,sob97}.

It has to be remarked that various aspects of the  somehow idealized 
presented scenarios,
 do not exclude each other but they can 
contribute to various stages of the reaction dynamics.
Moreover, the weak points of each explanation must not be neglected:
\begin{itemize}
\item 
{In applying the Rayleigh stability criteria, it has to be
realized that neck matter is not incompressible, but it can be warmed 
up and expanded.
The matching of the reaction
time at these energies, much shorter than in low energy fission, 
with the growth time-scale of Rayleigh instabilities \cite{bro90} could be 
also a problem.}
\item{
In a dynamical fission scenario, having in mind the
strong dissipation towards the scission point, 
the question raises about the possibility for the prefragment 
to have enough energy 
to escape with the velocity needed for reproducing 
the largest deviation from Viola systematics.}
\item{
The manifestation of a shearing off (sudden abrasion) process at 
lower energies
is certainly largely suppressed in a slower evolution towards
separation.}
\item{ Finally, the time scales for binary reactions at Fermi
energies may not be large enough to allow for a consistent 
statistical approach to dinuclear-like and  proximity adiabatic 
configurations, in fast continuous evolution
due to centrifugal motion of the $PLF$ and $TLF$ spectators.}
\end{itemize}

Therefore a unified and more systematic theoretical
investigation of the processes that develop in this energy range 
in semipheripheral collisions is required. It is certain that the intricate
interplay between effects related to mean-field and direct nucleon hard 
collisions
has to be properly described as well as the entrance channel
far from equilibrium regime. A reliable possible framework is provided by a
stochastic mean field microscopic 
approach. We consider a Boltzmann-Nordheim-Vlasov, ($BNV$), type 
transport model,
which guarantees a good description of the mean field dynamics,
we believe important for the physics in the Fermi energy range. It also
includes a collision integral term that consistently
accounts for the Pauli blocking. Moreover, the related fluctuations effects
are considered through a stochastic procedure, as detailed below.

Based on this approach, we
investigate how a fast fragment production mechanism
can raise  in semipheripheral
collisions at intermediate energies
and we try to characterize the most significant observables.
We explore the conditions that promote its manifestation
as well as the dependence on mean field properties 
and direct nucleon-nucleon collisions.
In this way we will be able to suggest the optimal experimental
selections (colliding ions, centrality, beam energies) for the observation
of this new dissipation reaction mechanism and also to extract
some fundamental information on the nuclear interaction in the medium.

We will strongly exploit the latter point in relation to
isospin effects in reactions with large charge asymmetry.
A peculiar attention will be payed to the transport properties 
of the isovector part of the equation of state ($asy-EOS$).

Several recent works were focused on the
role of isospin and charge equilibration in mid-pheripheral
colllisions around Fermi energies 
\cite{pog01}, \cite{row03}, \cite{sou03a}, \cite{sou03}, \cite{tsa03}.

We predict that some neck-observables 
are sensitive to the poorly known density dependence of symmetry
term of $EOS$ allowing to discriminate among 
various proposed parametrisations.
Therefore, very often in the following, our results
will be presented
as a comparison between different $asy-EOS$.
Moreover, once the isospin dynamics is better understood,
the fragment isospin composition can be considered as a useful
marker for the physical processes that take place at different
time-space regions during the reaction evolution.
We expect this physics to be largely produced in the coming years
at the new Radioactive Ion Beam ($RIB$) facilities in the Fermi
energy domain, e.g. see ref. \cite{EUR}.

In our quantitative evaluations we are focusing here  
mainly on the neutron rich, 
mass asymmetric
reaction $^{124}Sn+^{64}Ni$, at $35 AMeV$ bombarding energy,
but attention has been payed also to the corresponding neutron poor reaction 
$^{112}Sn+^{58}Ni$, for the reason which will become
clear below. These reactions have been experimentally 
studied within the $REVERSE$ experiment at $LNS$, Catania \cite{pag01}.
In spite of the low isospin densities reached with stable asymmetric
beams we will be able to see interesting symmetry effects in some 
selected observables very sensitive to the 
isospin dynamics. 

Based on these considerations we organize our work as follows.
In Section II the main ingredients of the considered
transport model are first briefly discussed.
Then we present global features of the reaction dynamics
for semipheripheral collisions.
The density contour plots allow to identity  several stages
of the collision and to extract the corresponding time scales.
A classification of the observed events is emerging. We identify a fast
$IMF$ production mechanism related to the neck dynamics 
which is generically called {\it  neck fragmentation}
and study its evolution with the impact parameter. 

In section III we survey the
kinematical properties of the $IMF$'s resulting from neck fragmentation
(Neck Originating Fragments, $NOF$),
including the velocity and angular distributions,  
for different $asy-EOS$ effective forces. 

In section IV, in order to better grasp
the physical process leading
to this mechanism,
we explore the influence of compressibility and
 nucleon-nucleon cross sections
on the neck dynamics.
We underline the essential role of volume instabilities
in the neck fragmentation process.

In section V we focus on the isospin dynamics in this mechanism
and point out that isotopic composition of $NOF$'s as well as the 
clearly evinced isoscaling parameters
are quite sensitive to the density dependence of the used $asy-EOS$. 

The emerging picture from our study and the main results are summarized 
in the Conclusions, section VI.

\section{SEMIPHERIPERAL COLLISIONS AROUND FERMI ENERGY: NECK FRAGMENTATION}

\subsection{Stochastic transport model - basic ingredients}

A new code for the solution of microscopic transport equations
of Boltzmann-Nordheim-Vlasov ($BNV$) type
has been written where the dynamics of fluctuations 
is included \cite{flu98}. 
The transport equations are solved following a test particle evolution on 
a lattice \cite{gua96,gre98}. In the collision term a parameterization 
of free $NN$ cross
sections is used, with energy and angular dependence.
The isospin effects on the nucleon cross section and
Pauli blocking are consistently evaluated.
The influence of in-medium reduced cross sections is studied in Sect.IV. 

We also adopt another approach to stochastic terms,
computationally much easier \cite{noise,col94}, 
based on the introduction of density fluctuations by a random sampling
of the phase space. The amplitude of the noise is
gauged to reproduce the dynamics of the most 
unstable modes \cite{col94}. 
For each system we have checked the equivalence of the two methods
in the description of the collision dynamics, from
fast particle emissions to the fragment production.

Our results are first discussed considering a ``soft''
effective interaction, corresponding to a compressibility
modulus $K=200 MeV$, see ref.\cite{iso02}. In the
section IV a hard $EOS$ parametrization with $K=380 MeV$ will be 
also considered.

Regarding the isovector part of the $EOS$, three different parametrizations 
for the density dependence of symmetry term are adopted,
the so called {\it asysoft, asystiff and superasystiff} $asy-EOS$, 
ref.\cite{iso02}.
We will refer to a "asystiff" $EOS$ when the 
potential symmetry term linearly increases with nuclear 
density, to a "asysoft" $EOS$ when the symmetry term shows 
saturation and eventually a decrease above normal density \cite{dit99},
and to a "asysuperstiff" behaviour if it has a
parabolic rise with density \cite{pra97,bli00}.

To achieve a reasonable statistics, for each impact parameter
and mean-field parameterization, 400 events were obtained
arriving at an overall total of more than 8000 events. 
The BNV transport code was run until freeze-out time, 
when the resulted fragments are quite far apart from each other,
under the action of only Coulomb repulsion.
In order to select the test particles
belonging to a given fragment a fast method, based on cuts in density,
was considered. We have also confronted it with other methods,
based on the interparticle relative distance criteria, obtaining
similar results. The mass, charge, angular momentum, quadrupole and
octupole deformations, excitation energies as well as
$CM$ positions and momenta for each fragment have been then evaluated.

Once the freeze-out space/momentum configuration has been fixed 
the Coulombian 
trajectories of all fragments were calculated
in order to generate the asymptotic angular and velocity distributions.

\subsection{Typical events and  time scales}

Our simulations indicate that from semicentral to peripheral collisions, 
above the impact parameter $b=4fm$, i.e. $b_{red} \equiv b/b_{max} \geq 0.37$,
 the reaction $^{124}Sn+^{64}Ni$, at $35 AMeV$,
has a mainly binary character. Based on density
contour plots, as well as from multiplicities 
and quadrupole/octupole fragment
deformations we can divide the
events at freeze-out in three main classes as follows:

a) binary events, with excited Target-like and Projectile-like
fragments ($TLF^*$, $PLF^*$) showing small deformations 
and therefore likely to remain so for long times; their 
sequential decay can be described by statistical models with reliable 
inputs for angular momentum and excitation energy;

b) binary events, but $PLF^*$
and/or $TLF^*$ acquire large quadrupole or/and octupole
deformation (especially $PLF^*$). These primary fragments now are 
expected to follow
a dynamically induced fission-like path, faster than a pure statistical one;

c) ternary events, with a $IMF$ directly emitted in less than 250-300 fm/c
from the reaction beginning.
The remaining $PLF^*$ and $TLF^*$'s are now in general less deformed.
At the considered energy, events with two or more $IMF$'s 
appear very rare (see the discussion of Section IV). 

\begin{figure}
\centering
\includegraphics*[scale=0.5]{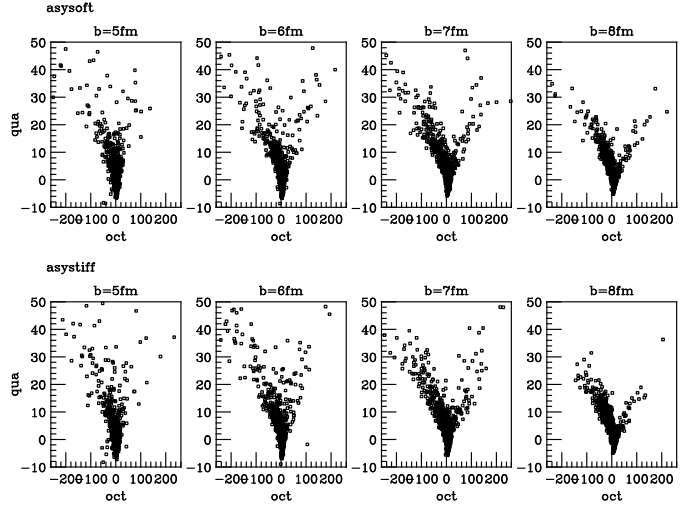}
\caption{Quadrupole ($qua$) versus octupole ($oct$) deformations of 
the fragments
 belonging to the event classes a) and b), see text.}
\label{quz1}
\end{figure}

As we have already mentioned the events of class b)
can also contribute to  the ``dynamical''
production of $IMF$'s we are studying here.  
We show in  Figure \ref{quz1}
the scatter plot of freeze-out quadrupole versus octupole moments,   
($qua=\sum_{i=1}^{A}(2z_i^2-y_i^2-x_i^2)/A$,
$oct=\sum_{i=1}^{A}[5z_i^3-3z_i (x_i^2+y_i^2+z_i^2)]/A$),
for all fragments produced in the events of classes a) and b),
for two asy-parameterizations.
The large dynamically induced  deformations
can drive especially the $PLF^*$'s towards a fast asymmetric 
fission. We recognize in the figure the two branches associated 
to $PLF^*$ (left) and to $TLF^*$ (right) respectively.
The corresponding signs suggest
pearshaped fragments oriented with the smaller deformation
towards the separation point. Unfortunately the mounting numerical
inaccuracy of the transport simulations cannot allow us to follow 
such events up to the scission point.

 Our approach will allow 
 a detailed and quantitative
analysis for only  the class c) of events. Nevertheless, based on the features
of this fast $IMF$  emission mechanism,
we will trace some conclusions also about the
dynamically induced fission expected to involve somehow longer time scales.

\begin{figure}
\centering
\includegraphics*[scale=0.5]{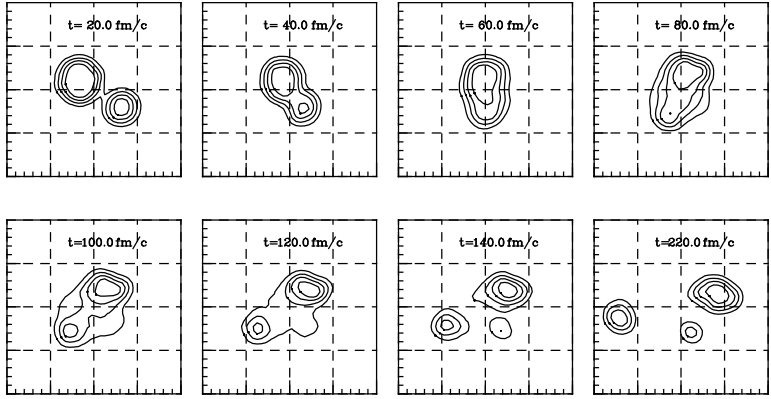}
\caption{Typical evolution of the density contour plot for a ternary event at 
$b=6fm$ for the reaction $^{124}Sn+^{64}Ni$ at $35 AMeV$.}
\label{massimden1}
\end{figure}

We show in Figure \ref{massimden1} the density contour plot, projected
on the reaction plane, 
of a typical ternary event belonging to the class c), for $b=6fm$.
Calculations are performed using the asy-stiff parameterization.
For the first $20-40fm/c$ from the touching time,
the two participants deeply interact
and some compression takes place. 
The system heats up and a relative expansion follows. 
In spite of its compact shape it still behaves as a
two-center object and we notice that a superimposed separate
motion of the $PL$ and $TL$ pre-fragments is effective. It induces
the formation of a neck-like structure
with a fast changing geometry, between $40fm/c$ and $140-160fm/c$, depending
on the impact parameter.

\begin{figure}[htb]
\centering
\includegraphics*[scale=0.5]{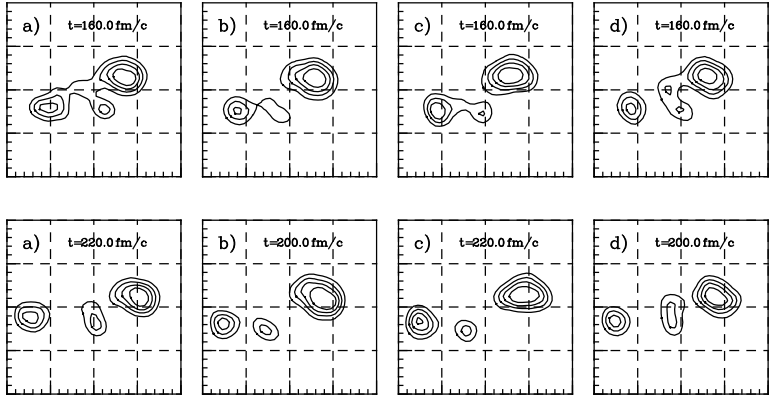}
\caption{Configurations corresponding to an early stage of fragment 
formation (top)
 and close to the freeze-out (bottom) for four ternary events a), b), c), d)
 observed in neck fragmentation
at $b=6fm$.}
\label{densample}
\end{figure}

This particular neck-instability dynamics favours the appearance of $IMF$'s,
after about $150fm/c$,
in a variety of places and ways as can be seen
by looking at Figure \ref{densample}. Here,
we have selected, for four events, two characteristic times,
the early phase of
fragment formation process and the configuration close to freeze-out.
We call the $IMFs$ produced by such a mechanism Neck Origination
Fragments ($NOF$'s).

\begin{figure}[htb]
\centering
\includegraphics*[scale=0.45]{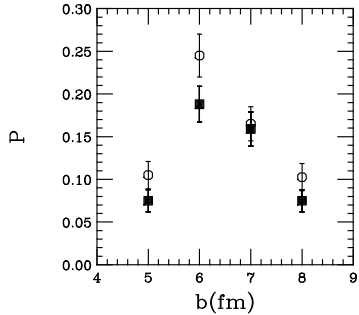}
\caption{The impact parameter dependence of the probability for ternary events.
White circles: neutron rich reaction. Black circles: neutron poor reaction. 
Asystiff $EOS$.}
\label{probimp}
\end{figure}

We can introduce the ternary event probability as the ratio,
 $N_{ternary}/N_{total}$, between the
number of events of class c) and the total number of events,
$N_{total} = 400$ for each impact parameter. 
It shows an interesting dependence on the
impact parameter, see  Fig. \ref{probimp}.
Its maximum, around  $25 \%$, is attained 
around mid-centrality, between $b=6-7fm$. The $NOF$ production probability
is then decreasing on both sides 
arriving at $10\%$ for $b=5fm$ and $b=8fm$. 
It becomes still smaller at $b=4fm$, in spite of
a stronger dissipation.
At variance, at larger impact parameters, $b=9fm$, a less overlapping 
and a  faster
 separation are also suppressing this mechanism.
The trend and the corresponding values at each impact parameter
are not sensibly influenced when other $asy-EOS$ are considered and therefore
we do not include the results in Fig. \ref{probimp}. We will show later
that the isospin diffusion, i.e. the isospin content of the $NOFs$,
is much more sensitive to the symmetry term of the used effective forces.

The probabilities are reduced of about $25\%$ for
the neutron poor reaction $^{112}$Sn+$^{58}$Ni, (black points).
We note that a similar difference in $IMF$ yields between neutron poor 
and neutron
rich systems was observed experimentally in multifragmentation events
for central collisions \cite{kun96}.
As we shall see in the Section IV, neck fragmentation
probabilities depend strongly on the
the nucleon-nucleon cross sections and compressibility.
All these results seem to indicate the relevance of volume instabilities
even for the dynamics of neck fragmentation.

\begin{figure}
\centering
\includegraphics*[scale=0.5]{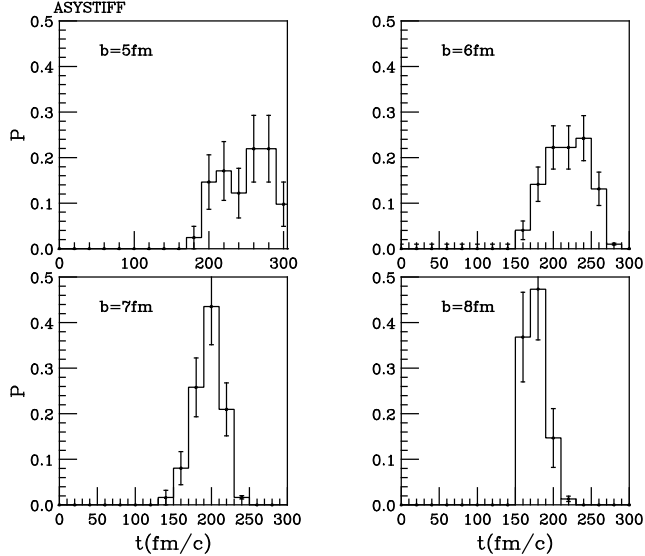}
\caption{$NOF$ production rate for impact parameters from $b=5fm$ to $8fm$ 
and asystiff $EOS$.} 
\label{prodrate}
\end{figure}

More insight on the $NOF$ production mechanism is gained looking at the 
time dependence of the
emission probability. At each $20 fm/c$ we can scrutinize the density 
distributions
searching for the appearance of new ternary events.
The corresponding emission probability is obtained as the ratio between 
the new ternary
events, identified within the specific time interval,
 and the total number of ternary events counted at freeze-out.
From Fig. \ref{prodrate} we see that the $NOF$'s production
is a fast process, taking place
between $120fm/c$ and $280fm/c$, depending on impact parameter.
We remind that in our initialization geometry the touching time is 
around $20fm/c$, see Fig.\ref{massimden1}.
One can notice that the distribution becomes sharper and peaked
at smaller times with increasing impact parameter, as expected.  

\begin{figure}
\centering
\includegraphics*[scale=0.55]{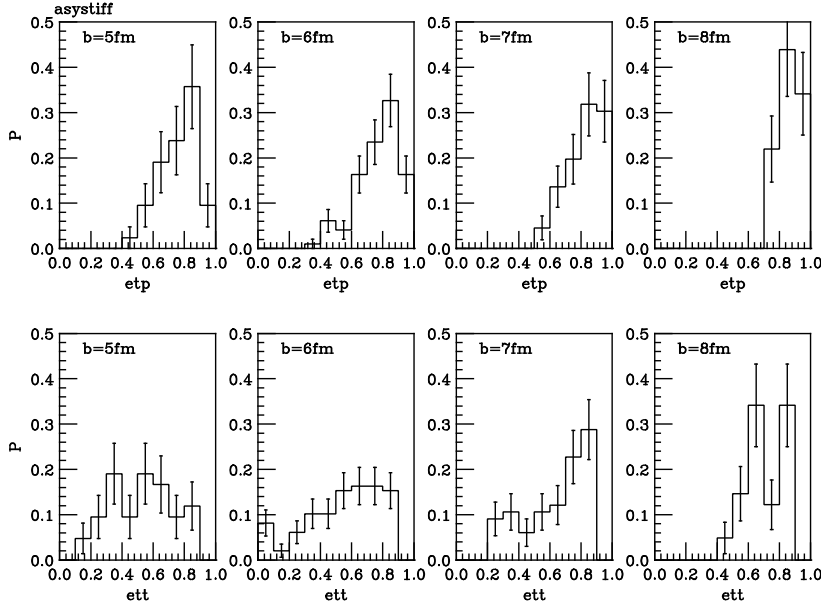}
\caption{Mass asymmetry $PLF-NOF$ (up) and $TLF-NOF$ (down) probability 
distribution for 
impact parameters from $b=5fm$ to $8fm$ and asystiff $EOS$. Freeze-out times.}
\label{etp}
\end{figure}

The primary $NOF$'s at the freeze-out time are characterized by 
reduced masses and charges reflecting specific spatial and temporal
constraints. The mass asymmetry 
distributions with respect to the $PLF$/$TLF$ are shown in Figure\ref{etp}.
We define $etp \equiv (M_{PLF}-M_{NOF})/(M_{PLF}+M_{NOF})$  and
$ett \equiv (M_{TLF}-M_{NOF})/(M_{TLF}+M_{NOF})$. As expected smaller $NOF$'s 
are in general produced
at larger impact parameters. We note that in this inverse kinematics reaction
the smaller target mass gives a wider mass asymmetry distribution for the
$TLF-NOF$ system. The mass asymmetry distribution of the fast produced $NOFs$
is much sharper with respect to the projectile-like fragments, with a very
clear separation between a Heavy $PLF$ and a light $NOF$.

\begin{figure}
\centering
\includegraphics*[scale=0.45]{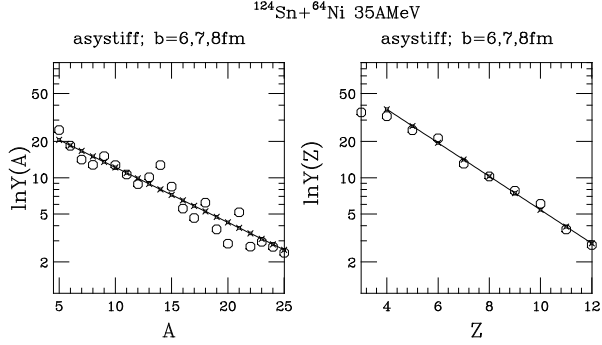}
\caption{Mass and charge distribution in neck fragmentation.} 
\label{semmult}
\end{figure}

The primary $NOF$'s yield actually displays
an  exponential dependence with
mass and charge, $Y(A) \sim exp(-0.12 A)$ and  $Y(Z) \sim exp(-0.32 Z)$ 
as is seen from the semilogarithmic plots in figure \ref{semmult},
where the contributions from each impact parameter are summed up with
the corresponding geometrical weight.
The values of slope parameters can depend on the constraints discussed before.

\begin{figure}
\centering
\includegraphics*[scale=0.5]{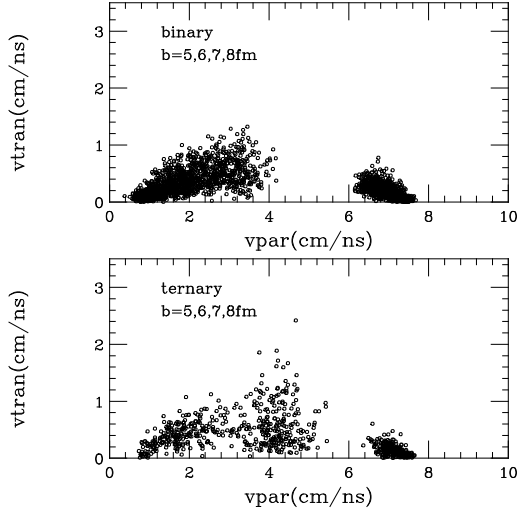}
\caption{Longitudinal and transversal velocity components of all 
fragments observed in
binary events (up) and ternary events (down).}
\label{vpartran2p}
\end{figure}

Finally for the same impact parameters the
asymptotic fragment velocity distributions, in the laboratory frame,
are plotted in Figure \ref{vpartran2p} both for binary and ternary events.
Here $'vpar'$ is the velocity component along the beam
direction and $'vtran'$ is the orthogonal part.
The $NOF$ are found in relatively wide midrapidity
region, close to the $CM$ velocity which is around $5.4 cm/ns$.

All these results guide us to individuate 
a fast neck break-up mechanism triggering the
formation of an $IMF$, localized in the midvelocity region, the $NOF$.
The emission take place in a temporal window
between $140fm/c$ to $260fm/c$ from the reaction beginning.
The best physical conditions
are created at intermediate impact parameters, between $b=5fm$ and $8fm$ for
the considered system. This is what we call {\it neck fragmentation mechanism}.
Our task in the following sections
is to study its particularities 
performing a more detailed analysis of $NOF$'s velocities and 
angular distributions as well
as of their isospin content.

\section{NECK FRAGMENTATION: KINEMATIC CORRELATIONS}

In order to reveal the nonstatistical features of the $NOF$ production
we will look here at some clear kinematic correlations of dynamic
nature. The corresponding observables can be promptly measured in 
exclusive experiments.

\subsection{Deviations from Viola systematics}

\begin{figure}
\centering
\includegraphics*[scale=0.55]{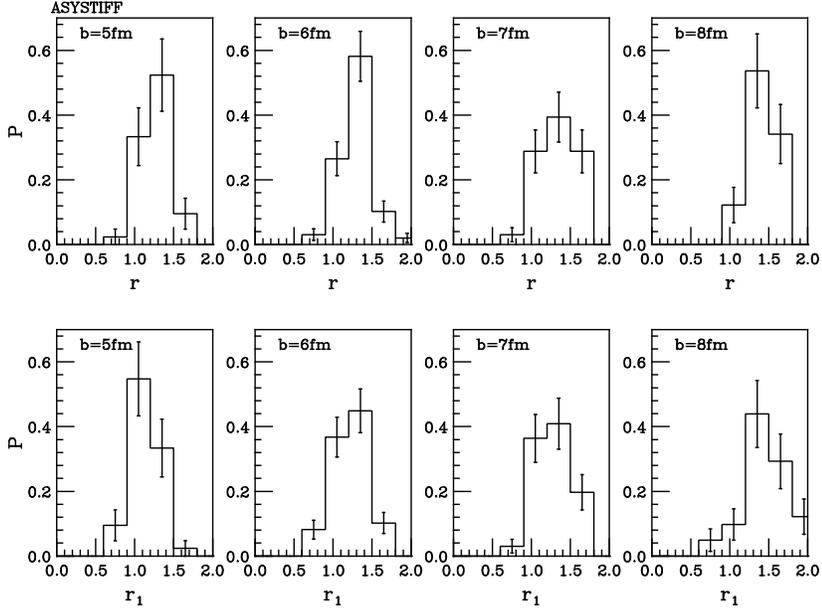}
\caption{Probability distributions of the deviations from Viola 
systematics with respect to $PLF~(r)$ (top) and $TLF~(r1)$ (bottom)
respectively. Impact parameters 
from $b=5fm$ to $8fm$ and asystiff $EOS$.} 
\label{hist0vio2}
\end{figure}

We first construct the asymptotic relative velocitities of 
the neck-produced $IMFs$ with respect to the  $PLF$ ($TLF$),
$v_{rel}(PLF,TLF) \equiv {\vert {\bf v_{PLF,TLF}} - {\bf v_{NOF}} \vert}$.
We compare these quantities
to the relative velocities from a pure Coulombian driven separation, 
in a hypothetical statistical fission process of a compound $PL$ or $TL$
system, 
as provided by the Viola systematics \cite{vio85,hin87}:
\begin{equation}
 v_{viola}(1,2)= \sqrt{\frac{2}{M_{red}}(0.755 \frac{Z_{1}Z_{2}}
{A_{1}^{1/3}+A_{2}^{1/3}}+7.3)}
\end{equation}
 were $A_{1},A_{2},Z_{1},Z_{2}$ are the mass and charge number of 
the fission products
and $M_{red}$ is the corresponding reduced mass.
We introduce the quantities $r(r1)$, as the ratio between the observed
$PLF(TLF)-NOF$ relative velocity and the one obtained from Viola systematics, 
 i.e.
$r=v_{rel}(PLF)/v_{viola}(PLF)$, ($r1=v_{rel}(TLF)/v_{viola}(TLF)$).
For each impact parameter, we calculate  the  $r$ and $r1$
probability distributions as shown in  Figure \ref{hist0vio2}.
Most of them have relative velocities with respect
to $PLF$ ($TLF$) from $25\%$ to $70\%$ larger than the 
values provided by the Viola systematics.
The distributions move towards larger deviations
with increasing impact parameter.

\begin{figure}
\centering
\includegraphics*[scale=0.45]{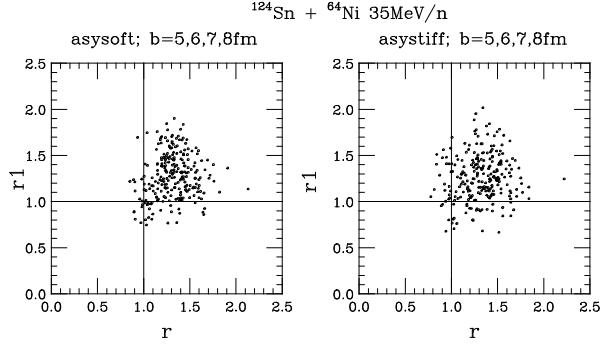}
\caption{Wilczynski-2 Plot: correlation between deviations from Viola 
systematics (see the text for definitions and ref.\cite{wilcz}). Results 
are shown for two $asy-EOS$.}
\label{massimvio2}
\end{figure}

It can by argued that a fragment
showing a large deviation from Viola systematics with respect to the $PLF$ is
actually more correlated to the $TLF$, i.e. to a large $r$ 
will correspond a closer to one
$r1$ and viceversa, a situation that will resemble more
to a statistical fission scenario rather than to a dynamical production.
We have tested this possibility plotting $r1$ against $r$ for each $NOF$,
in  Figure \ref{massimvio2}, \cite{wilcz}. 
The solid lines represent the locuses of the $PL$- ($r=1$) and $TL$-fission
($r1=1$) events respectively.
The observed values $(r,r1)$ mostly appear simultaneously 
larger than one suggesting a weakened $NOF$ correlation with {\it both} 
$PLF$ and $TLF$, ruling out the statistical fission mechanism.

In some respect the process is more displaying an analogy
with the participant-spectator scenario. However the dynamics appear much 
richer than in this simple sudden abrasion model, where the locus of
the $r-r1$ correlation should be on the bisectrix, apart the Goldhaber
widths, see ref.\cite{luk03}. Here we see wide distributions revealing 
a broad range of
fragment velocities, typical of the instability evolutions in the neck region
that will lead to large dynamical fluctuations on $NOF$ properties.

An induced asymmetric fission
can also manifest some deviations from Viola systematics.
In general, from our simulations, this mechanism takes place 
on longer time scales
comparing to the neck fragmentation. In any case if a light fragment escapes
later from its $PL/TL$ partner, the friction certainly will attenuate 
the dynamical
effects and relative velocities will deviate less from Coulomb contribution.
In the plane $r-r1$ such events are located closer to the line
$r=1$ or $r1=1$. Moreover a prolonged $PL$ or $TL$ joint propagation will 
determine also different
angular distributions in comparison to the neck fragmentation. This will
be then another interesting correlation to look at. See the next point.

\subsection{Narrow angular correlations around alignement}
 
In the situation of a dynamically induced fission 
the reaction is expected to proceed as a two step process.
Two primary excited fragments after a deep inelastic interaction, $PLF^{*}$ and
$TLF^{*}$ are first produced. Their separation axis has  
direction and orientation given by the $PLF^{*}$-$TLF^{*}$ relative velocity,
${\bf n}_{S}=({\bf v}_{PLF^{*}}-{\bf v}_{TLF^{*}})$.
The reaction plane is then defined by the normal
vector constructed from
the beam axis and separation axis as 
${\bf n}=({\bf n}_{S} \times {\bf n}_{B})$,
where  ${\bf n}_{B} ={\bf v}_{CM}$.
In this plane we introduce the $z$-axis along the
beam direction and with the same orientation and the $x$-axis orthogonal 
to it and oriented 
from target to projectile.

In the second stage, depending on their excitation energy and deformations, 
 the primary
fragments can decay into the fission channel. 
Since in our case the projectile is heavier and more fissile we will
consider the case of a $PLF$ induced fission. The fission axis is determined
by the relative velocity of fission products,
denoted here as $H$ (Heavy) and $L$ (Light), to suggest an asymmetric process
of interest for the comparison with our $IMF$ production:
 ${\bf n}_{F}=({\bf v}_{H}-{\bf v}_{L})$. 
The in-plane azimuthal angle, $\Phi_{plane}$, is the angle between
the projection of the fission axis onto the reaction plane and 
the separation axis.
We follow the convention introduced in \cite{ste95}, and consider
 the angle $\Phi_{plane}$
positive when ${\bf n}_{S} \times {\bf n}_{F}$ and ${\bf n}$ both point 
into the 
same half-space. A $\vert \Phi_{plane} \vert \simeq 0$ collects events
corresponding to asymmetric fissions of the $PL$-system very aligned
along the outgoing $PL-TL$ separation axis. A statistical fission dominance 
should correspond to a flat $\Phi_{plane}$ distribution.

\begin{figure}
\centering
\includegraphics*[scale=0.55]{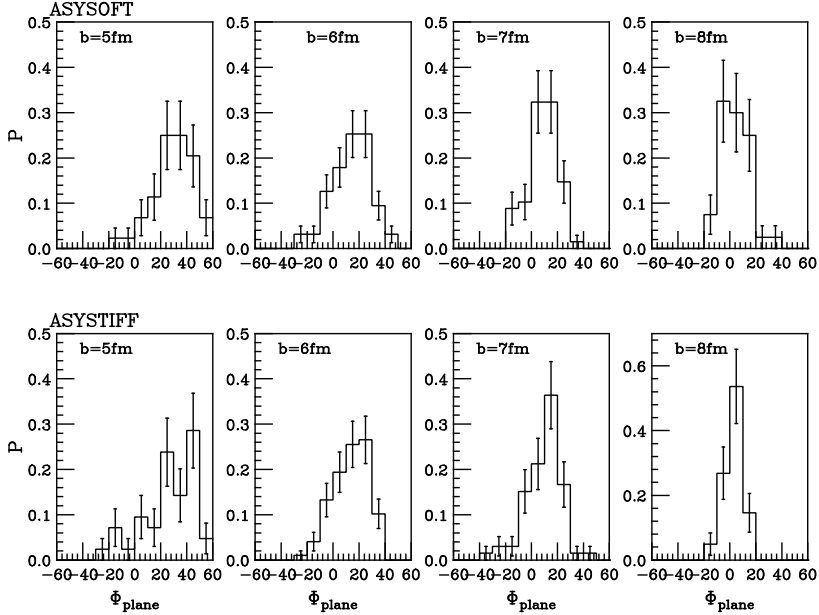}
\caption{The $\Phi_{plane}$ probability distributions: top, asysoft EOS, 
bottom,
 asystiff EOS. Impact parameter range from $b=5fm$ to $8fm$.} 
\label{phiplan}
\end{figure}

It is then instructive, as we will see, to adopt the same language and 
system of definitions when studying
the angular correlations in the neck fragmentation mechanism.
The $\Phi_{plane}$ distributions, corresponding to the same $NOF$
events analysed before, are shown in figure
\ref{phiplan} for impact parameters from $b=5fm$ to $8fm$
and two $asy-EOS$, asysoft and asystiff.
The $NOF$'s are observed in a quite
limited angular window,
close to full alignment configurations,
$\Phi_{plane}=0^{\circ}$. The range of values 
$\Delta \Phi_{plane} \approx 80^{\circ}$ for $b=5fm$ is
becoming narrower with increasing
impact parameter.
However the angular distributions are not symmetrically distributed
around  $\Phi_{plane}=0^{\circ}$, but shifted toward positive values.
This behavior is more evident at smaller impact parameters but even
at $b=8fm$ for more than $60-65\%$ of ternary events 
$\Phi_{plane}$ is positive.
These properties reflect entrance channel effects and early stages 
of the neck fragmentation dynamics and $NOF$ formation.

\begin{figure}
\centering
\includegraphics*[scale=0.5]{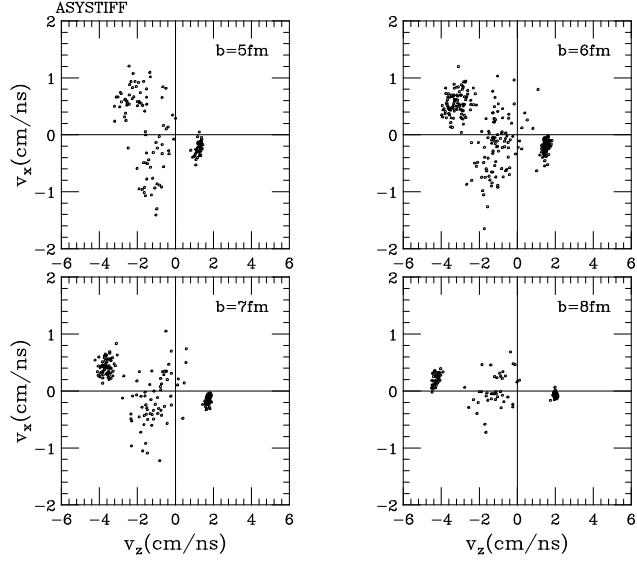}
\caption{Center of mass $z,x$-velocity components distributions.} 
\label{velcm2}
\end{figure}

Indeed, this can be clearly realized from the figure
\ref{velcm2} were we plot the freeze-out $CM$ velocity components
$v_z$ and  $v_x$, of all fragments identified in the ternary events.
We can easily localize the $PLF$'s and $TLF$'s positions.
Their $x$-component
velocities reflect the orbiting motion before separation 
(see Figs.\ref{massimden1}, \ref{densample}).
The freeze-out $v_z$'s, when compared to the corresponding initial velocities
in the $CM$, are an indication of the the kinetic energy dissipation.

Most of the $NOF$'s have their $CM$ velocity components $v_z$, $v_x$ both
negative. This a consequence of their early decoupling combined
to the Coulomb repulsion. A simple analysis upon these velocity distributions
indicates that the corresponding $\Phi_{plane}$ angle has to be quite
small and positive as it was seen in the previous figure.
The fast neck-fragmentation mechanism is essential to understand this
behavior, in particular if we look at the third quadrant points
which are well off the rotational axis connecting the $TL$ and
$PL$ velocity regions.
Here if only Coulombian effects are manifest
$v_x$  has to be anticorrelated with  $v_z$:
if moving counterclockwise,
$v_x$ should decrease in absolute value meanwhile the modulus of
$v_z$ will increase. We do not see this correlation: rather they look
proportional. This can be associated with the collective expansion 
of the neck matter
on the way toward separation, between $60 fm/c$
and $160 fm/c$, which, because of geometrical configuration
(see also Fig.\ref{massimden1}) will enhance
mainly the negative $x$-component of the fragment velocity.
 
\begin{figure}
\centering
\includegraphics*[scale=0.5]{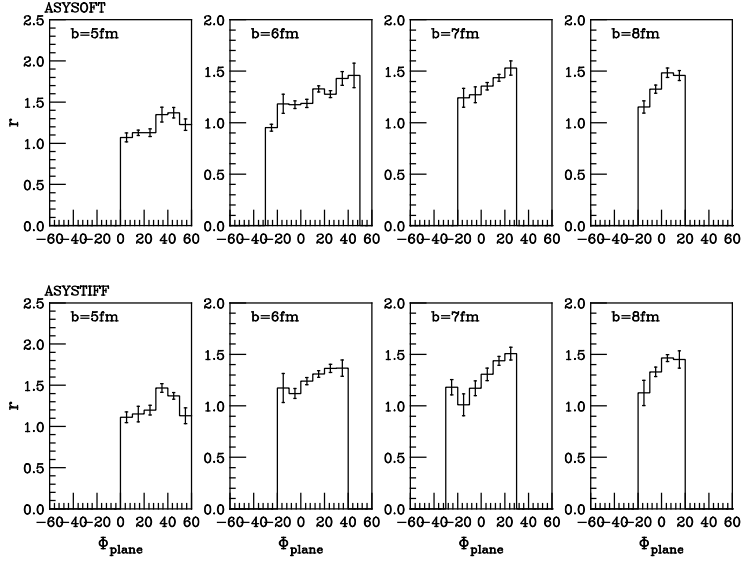}
\caption{$\Phi_{plane}$ angle dependence of the deviation from 
Viola systematics.} 
\label{hist0vethe}
\end{figure}

 This is a new dynamical effect which adds its contribution to the observed
deviations from Viola systematics.
Indeed, a $NOF$, depending on the time and
place where is formed, will experience in a certain degree 
this collective motion. From Fig.\ref{velcm2} we can see that for $ b=5,6,7fm$
($v_z$,$v_x$) can reach values up to ($-2cm/ns$,$-1cm/ns$). 
The $IMF$s with largest negative $v_x$ components (and largest 
positive $\Phi_{plane}$ angles) represent the earliest fragment formation
in the neck region: they should show the lowest correlation to the
$PL$ residues. 
If this is
true then we expect that the deviation from Viola systematics
has a rising trend when moving toward positive  $\Phi_{plane}$ angles.
The value of $\Phi_{plane}$ for which this maximum
is attained as well as corresponding value of maximum
are directly related to the dynamical effect just discussed.
Beside this we have to add that, less frequently in this mass asymmetric
system, the fast $NOF$s can be formed also  
on the other side of the neck and so they will experience an 
enhancement of $v_x$
in positive direction. Since in this case $\Phi_{plane}$ is negative,
we expect some rise of the $Viola-violation$ $r-$parameter, more discrete, 
also at negative angles.
This features are indeed observed when we look at the correlation
with the Projectile-like system
$\Phi_{plane} -r$ in figure \ref{hist0vethe}.
We note that this mechanism naturally predicts a dip in the $r-$values
in correspondence to the $\vert \Phi_{plane} \vert \simeq 0$ region,
i.e. for $PLF$ breakings fully aligned along the separation axis.

When the light fragment remains attached longer to the 
heavy one, the $\Phi_{plane}$ ditribution becomes flatter and 
the deviation from Viola systematics 
diminishes gradually.  However, as mentioned before, it is hard
to follow this process (the dynamical fission) within BUU-like
approaches. 



Summarizing this discussion, we conclude that the $\Phi_{plane}$
angle is a very appropriate observable to disentangle among various mechanisms.
For the asymmetric reaction considered here, 
an anisotropic distribution, populating preferentially 
positive small angles, but
extended also at negative angles,
characterize neck fragmentation mechanism. 
The largest deviations from Viola systematics due to neck expansion 
will be reached
at positive angles.
 An induced dynamical fission, expected as the next fast process in a temporal
ordering, will be characterized by a wider but still
anisotropic distribution  
and by smaller deviations from Viola systematics.
Finally, closer to a statistical fission, large angles,
in absolute values, can also be reached. Meanwhile more events with
mass symmetric $PLF$ fission will appear. The Heavy/Light mass ratio
$A_H/A_L$ appears then also a good parameter to select fast neck-dynamics
effects.

\section{UNVEILING THE NECK FRAGMENTATION MECHANISM: 
ROLE OF VOLUME INSTABILITIES}

The results presented in the previous sections were obtained
considering a free, energy dependent nucleon-nucleon cross sections
and a soft $EOS$ parametrization with a compressibility
modulus around 200 MeV.

In the Fermi energy domain the mean-field and the two-body collisions
have comparable effects on the dynamical evolution.
A classical example is provided by the 
collective flows: none of the two ingredients, alone, can 
reproduce the observed experimental behaviour.
Here we are concerned with their influence on the neck fragmentation dynamics.

\subsection{EOS dependence}

\begin{figure}
\centering
\includegraphics*[scale=0.5]{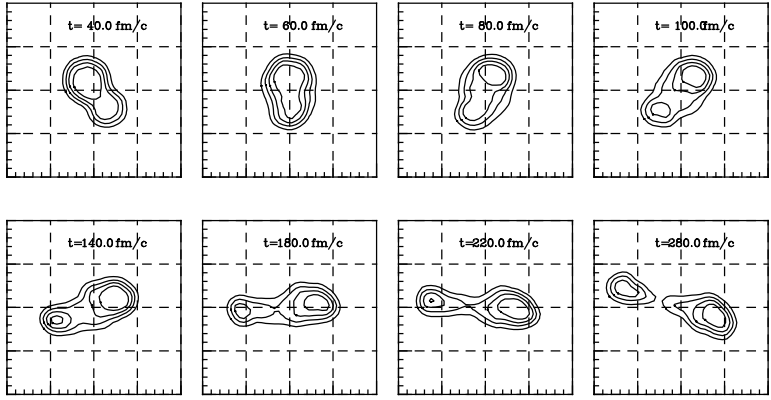}
\caption{Typical evolution of density contour plot for a $b=6fm$ event
simulation with a hard $EOS$.}
\label{denhard}
\end{figure}

We first perform new simulations adopting a hard, ($K=~380MeV$), equation of state 
and
keeping unchanged the nucleon-nucleon cross-sections.
We have run $400$ events at a chosen impact parameter $b=6fm$, for which,
with the soft mean field the neck fragmentation had the most frequent 
appearance. During the first $300fm/c$ no ternary
event was observed. From density contour plots, like Figure \ref{denhard},
a quite different evolution is apparent.

A more consistent neck will bind PL and TL prefragments
longer time together. Moreover, no evident decoupling of the
participant region is manifest. Rather, the separation takes
place by the rupture of a very elongated neck structure.

\begin{figure}
\centering
\includegraphics*[scale=0.45]{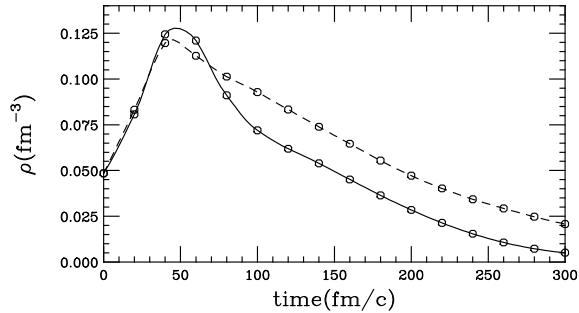}
\caption{Density evolution in the neck region (see the text) 
for soft $EOS$ (K=200MeV, solid) and
stiff $EOS$ (K=380MeV, dashed).} 
\label{densevol}
\end{figure}

This new reaction evolution must be related to the different
bulk compression-expansion dynamics. 
In order to clarify this point, in Fig.\ref{densevol}
 we plot the time evolution of the average nuclear
matter density in a cubic volume of $10fm$ side,
centered in the $CM$, in the two cases.
With a soft $EOS$ (solid line) we see a large initial density oscillation.
The neck zone will deeply enter the low density spinodal instability region
with a fast cluster formation on a time scale comparable with the
$PLF-TLF$ separation time due to the rotation. This will lead to
the observed early $NOF$ decoupling from moving $PL$ and $TL$ prefragments.
The growth time of density instabilities is
also in agreement with the observation that fragments appear after $140 fmc/c$
\cite{bar98}.

With a hard $EOS$ (dashed line) the density oscillation is much reduced,
the neck-system is hardly entering the dilute unstable region. Instabilites,
if any, will grow on a very long time scale and the neck will break before
any $PLF/TLF$-decoupled cluster formation. 

The geometrical and temporal constraints will limit the size of the 
formed $NOF$'s,
setting an important difference with respect to the spinodal decomposition
in multifragmentation processes, \cite{col94a}, \cite{gua96},
where it was involved
as the kinetic  process that initiates the fragment formation.

Nevertheless the surface fluctuations in the neck region may
play also a role in causing the density inhomogenities. Our results seem 
to show that the volume contributions are essential and
$NOF$ multiplicities will be sensitive to the $EOS$ stiffness.

\subsection{Influence of the $NN$ cross sections}

\begin{figure}
\centering
\includegraphics*[scale=0.55]{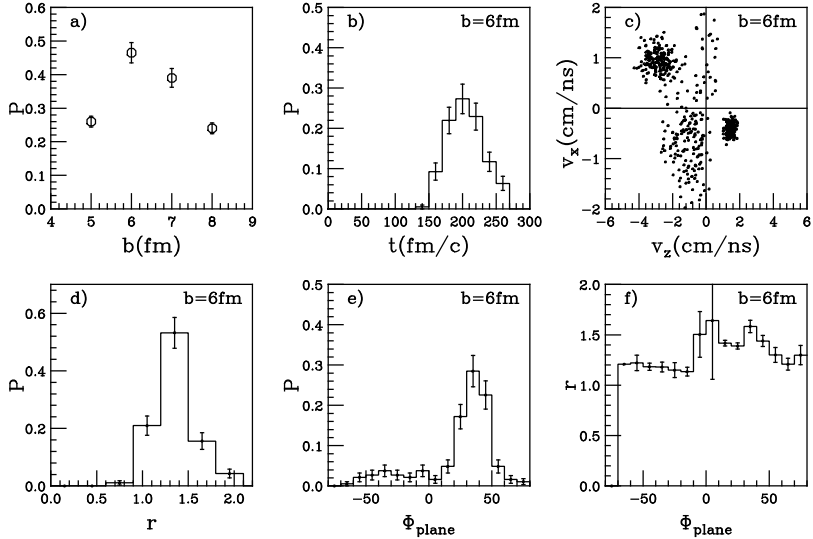}
\caption{Main features of neck fragmentation simulations with reduced 
$NN$ cross sections (see text).} 
\label{sigmahalf}
\end{figure}

Two-body collisions prove also to be important in the dynamics
of this mechanism. We have performed a new set of calculations, with a 
soft $EOS$
but reducing to a half the values of nucleon-nucleon cross sections.
Quite puzzling, the number of ternary events is increasing by a
factor two over all the range of impact parameters, see 
Figure \ref{sigmahalf} a).
We even noticed events with emission of two $NOF$s, around $10\%$ 
of the total.

The $NOF$ formation is somehow anticipated. We observe earlier high 
production rates,
see for comparison, the Figures \ref{sigmahalf} b) and \ref{prodrate}.
The highest rates, for $b=6fm$, are now attained between 
$180fm/c$ and $230fm/c$ meanwhile
previously the maximum was between $200fm/c$ and $250fm/c$.  

The $NOF$'s $v_z$ velocity components are not much affected by the cross 
sections reduction while a
definite influence is observed on the transversal 
components $v_x$, which
now, for $b=6fm$, extend from $-2cm/ns$ to $2cm/ns$, as can be seen from
 Fig. \ref{sigmahalf} c). This behaviour, in conjuction to the fact 
that $PLF$ velocitities
are less damped, has as a direct consequence some larger deviations
from Viola systematics, see Fig.  \ref{sigmahalf} d), and a shift of the
$\Phi_{plane}$ distribution toward larger positive values,  
Fig. \ref{sigmahalf} e).

We remind that the collisional friction is essential for the 
midrapidity stopping power, the warm and dense neck zone formation
and the dynamical fluctuations which originate from the instabilities.
The presented results show that the tuning of the collisional
contribution also modifies the neck separation dynamics. Reduced 
$NN$ cross sections will favour the instabilities development. 
Moreover, owing to a smaller stopping, the neck-breaking
time is better matching the time scales of the instabilities that lead
to a $NOF$ formation. The characteristics of the process, $NOF$'s emission
probability, angular distributions and deviations from Viola systematics,
 are consequently clearly affected.

\section{THE ISOSPIN DYNAMICS IN NECK FRAGMENTATION}

From the results shown in the previous sections we can notice that
the symmetry term of $EOS$ does not influence sensitively the main pattern
of the neck fragmentation mechanism:
probability of ternary events, deviations from Viola 
systematics and $NOF$ angular distributions.
We have to say that with the studied system we cannot reach
high charge asymmetries, the asymmetry parameter $I \equiv (N-Z)/A$
ranging from $0.193$ for the projectile and $0.125$ for the target
to an average $I=0.17$. 

However we will show in this section that we can clearly select other 
quantities, as average isospin 
content of fragments
or isotopic probability distributions,
that appear much more sensitive to the symmetry energy term.
In other words observables that are widely related to the
isospin diffusion during the neck dynamics appear more sensitive
to transport properties of the effective interaction in the
isovector channel.

On the other hand we have concluded that neck fragmentation is a quite
fast mechanism, driven by dynamical effects in the first $150fm/c$-
$300fm/c$, depending on the impact parameter.
Therefore it can provide distinct conditions for studying
the isospin dynamics on various time scale.
Moreover, due to density variations in the
neck regions, we hope that in this way we could directly test the 
density dependence
of the symmetry potential.

\subsection{Isospin dynamics}

\begin{figure}
\centering
\includegraphics*[scale=0.5]{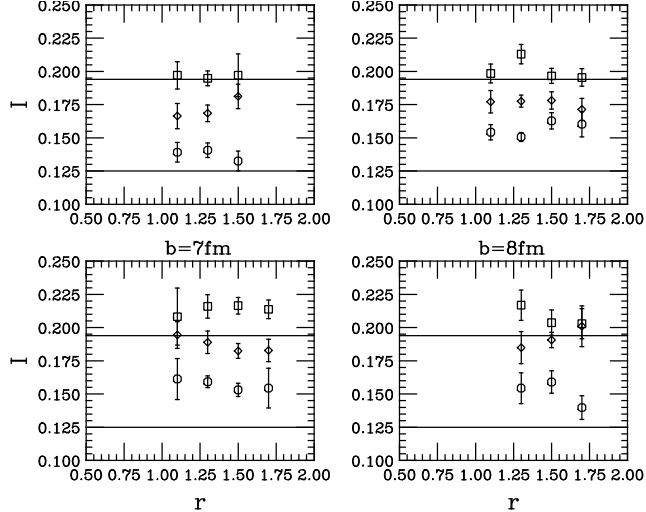}
\caption{$NOF$ isospin content for $asysoft$ (circles), $asystiff$ (rombs) 
and $superasystiff$ (squares)
$EOS$ as a function of $r$-deviation from Viola systematics, 
at impact parameters from $b=5fm$ to $8fm$. The two solid lines
represent the mean asymmetries of the projectile (top) and target (bottom). }
\label{papisorr}
\end{figure}

For the three $asy-EOS$ introduced in Section 2 we plot in 
the Fig.\ref{papisorr}
the average isotopic composition $I$ of the $NOF$'s
as a function of the $PLF$ $r$-deviation from Viola
systematics. At all impact parameters clear differences 
are evidenced. The average asymmetry does not depend
strongly on $r$. We notice however that it increases proportionally
with the slope of the symmetry potential as a function of
density around and below normal density. The {\it superasystiff}
parametrization, i.e. with an almost parabolic increasing behavior
around $\rho_0$, produces sistematically more neutron rich
$NOF$s.

\begin{figure}
\centering
\includegraphics*[scale=0.45]{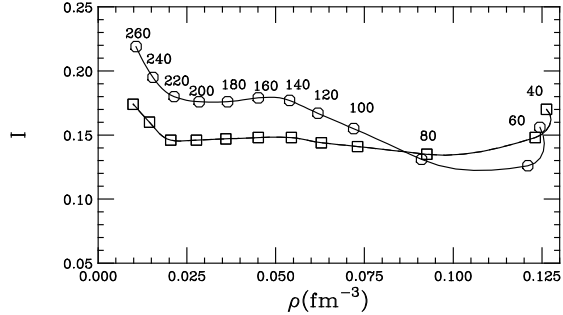}
\caption{Mean asymmetry-density correlations in a cubic volume
centered around the $CM$ during the reaction evolution, see the text.
The points are  obtained by averaging over all events
for asysoft (squares) and superasystiff (circles) $EOS$ at $b=6fm$.}
\label{denasycor}
\end{figure}

This effect is clearly due to a different neutron/proton migration
at the interface between $PL/TL$ ``spectator'' zone around normal
density and the dilute neck region where the $NOF$s are formed.
In order to understand the details of this isospin diffusion dynamics we follow
the time evolution of the average density and average asymmetry 
in the spatial cubic volume of $10fm$ side around the $CM$ introduced in the
previous section and containing mainly the neck zone. The corresponding
correlation is shown in Fig.\ref{denasycor} for asysoft and superasystiff
parametrizations. 
The numbers close to the each symbol indicate the corresponding
time step. 

The density variations in the considered volume
are very similar for the two $asy-EOS$, at each time the circle and square 
points are aligned roughly at the same mean density $\rho$. This is not 
surprising since the density oscillations are due to the stiffness of the 
symmetric (isoscalar) part of the $EOS$ which is kept exacly the same
(soft choice here) for the two different $asy-EOS$ parametrizations.
Actually the overall compressibility in asymmetric matter is modified
by the density dependence of the symmetry term, see \cite{dit99}, but
the effect is not large and proportional to the square of the 
mean asymmetry $I$. This can explain the very tiny alignement shift
between circles and squares in Fig.\ref{denasycor}. 

The initial average asymmetries  have  values
close to that of the composite system ($I=0.17$). The differences  
until  $80fm/c$ 
can be assigned to a different relative fast neutron to proton   
emission from the central region. In this time interval
we have a density oscillation, see Fig.\ref{densevol}, and so
neutrons are probing some more mean field repulsion in the superasystiff
case (the circles slightly below the squares). 

After $80fm/c$ the dilute neck region is forming, see Fig.\ref{massimden1},
and a clear dependence on the used $asy-EOS$
of the isospin dynamics is evident.
A much larger neutron enrichment of the neck zone for the
asysuperstiff $EOS$ takes place. We can understand this results
in simple terms. In presence of
density and isospin gradients, as in the case of neck fragmentation, the nucleons will
feel the superposed effects of the forces due to the isoscalar and isovector
part of the mean field. In the asysuperstiff case
the neutrons will be accelerated towards the neck region since the 
repulsion of 
the symmetry part of the mean field is sharply increasing around
normal density. This migration from spectator to neck at the interface
compensates in part the neutrons that moves out from the participant region
in presence of the expansion causing finally the observed enrichment.
This effect is largely reduced in the asysoft case due to the flatness
of the density dependence of the symmetry mean field around $\rho_0$.
 
This trend is not modified when reducing by half the nucleon-nucleon 
cross-sections. The reason is related to the fact that at these energies, 
 and lower densities, we are concerned with a transport regime intermediate 
between ballistic and hydrodynamical.

From the last two figures we see that the isospin content of $NOF$s
 carries important
information on the isovector part of the effective interaction and
the related isospin dynamics in the early stages of
the reaction. We expect a different isospin pattern for
the $IMF$'s resulting from induced and/or statistical fission of the
more charge symmetric $PLf^*$'s and $TLF^*$'s.  

\subsection{Isoscaling}

In recent years interesting isospin effects on fragment production
were evidenced through a isoscaling behaviour, observed
first in multifragmentation \cite{xu00,tsa01}. 
It was experimentally observed that
when compared two different reactions, 
one neutron rich (label $2$) and the one neutron poor (label $1$), 
the ratio between the
corresponding yields of a given $N,Z$ isotope $R_{21}=Y_2(N,Z)/Y_1(N,Z)$
verifies the following relation:

\begin{equation}
\ln R_{21}~=~Cost + N \alpha + Z \beta
\end{equation}
with $\alpha$ and $\beta$ isoscaling parameters.
In a grand-canonical statistical approach these parameters are
related to the neutron (proton) chemical potential differences
between the nuclear environments where the fragments are  
created in the two reactions:
\begin{equation}
~~~~~\alpha~\equiv~\frac{\Delta \mu_n}{T}~~~~,
~~~~\beta~\equiv~\frac{\Delta \mu_p}{T}~~~~~,
\end{equation}
In turn, these are related to the symmetry energy properties:
\begin{eqnarray}\label{eqsym}
\Delta \mu_n~=~\rho \frac{\partial E_{sym}}{\partial\rho} (I_2^2 - I_1^2)
 + 2 E_{sym} [(I_2-I_1) - \frac{(I_2^2 - I_1^2)}{2}] \nonumber \\
\Delta \mu_p~=~\rho \frac{\partial E_{sym}}{\partial\rho} (I_2^2 - I_1^2)
 - 2 E_{sym} [(I_2-I_1) + \frac{(I_2^2 - I_1^2)}{2}] \nonumber 
\end{eqnarray}
where $ E_{sym}$ is defined by
\begin{equation}
E(\rho,I) \equiv \frac{\epsilon(\rho  ,I)}{\rho  }
 = E(\rho) + E_{sym}(\rho  ) I^2 + O(I^4) +...
\end{equation}











$~~~\vert {\Delta \mu_p} \vert < \vert {\Delta \mu_n} \vert$.
 

Since  always we have the following relation:
\begin{equation}
\ln \Big(\frac{N_2/Z_2}{N_1/Z_1}\Big)~\equiv~\alpha-\beta~=
~\frac{4}{T} E_{sym}(\rho) (I_2 - I_1)
\label{equil}
\end{equation}

a large interest is rising on the possibility of a direct measurement
of the symmetry energy in the fragment source from the isoscaling
$\alpha,\beta$ parameters.

We have tested if the isoscaling behavior can manifest
in neck fragmentation too, when characteristic time scales
are shorter and several dynamical features will make
questionable the previous statistical equilibrium arguments.
To this purpose we have performed the same calculations
for the neutron-poor $^{112}Sn+^{58}Ni$, accumulating the
same ``statistics'', $400$ events for each $b=6,7,8fm$
impact parameter.

\begin{figure}
\centering
\includegraphics*[scale=0.5]{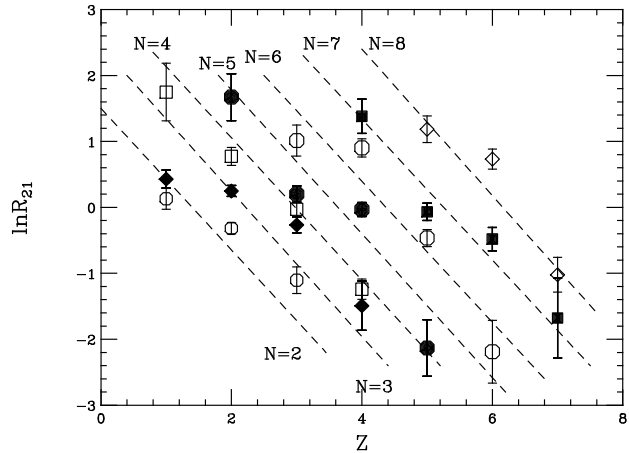}
\caption{Isoscaling in neck fragmentation: $lnR_{21}$ dependence on Z.} 
\label{isobeta}
\end{figure}

In figures \ref{isoalpha}, (\ref{isobeta}) are  shown the $N$ ($Z$)
dependence of $lnR_{21}$, for $Z=1$ to $Z=9$ light fragments
produced in the neck region, as obtained from our calculations
for a $asystiff-EOS$ parametrisation:

\begin{figure}
\centering
\includegraphics*[scale=0.5]{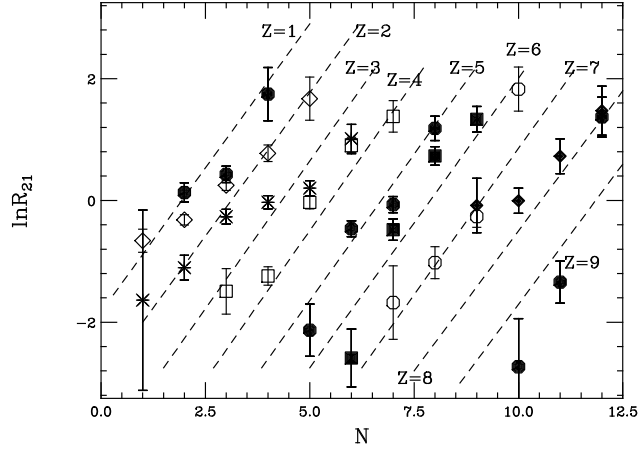}
\caption{Isoscaling in neck fragmentation: $lnR_{21}$ dependence on N.} 
\label{isoalpha}
\end{figure}

We clearly see a nice isoscaling signal, exponential $N$- and $Z$-
dependence of the yield ratios with very well defined $\alpha,\beta$
slopes. All that in spite of the dynamical features of these $NOF$s
that we have extensively described in the previous sections. 

Although we cannot use explicit equilibrium relations, like Eq.\ref{equil},
we still expect a symmetry energy dependence of the isoscaling
parameters.
Since the fragment formation takes place in the neck region,
we can predict that its isospin content will dictate the values
of the isoscaling parameters.
Indeed we may assume that for a
neutron poor system, closer to symmetric case, the differences
between various $asy-EOS$ on the isotopic and isotonic distributions
are reduced, and in a first approximation identical. 
At variance, for neutron rich
system, by passing from $asysoft$- to $superasystiff$-$EOS$, we have seen
that a more neutron-rich
neck region is forming, favorating a relative larger production of 
more asymmetric $IMF$'s. Therefore, the corresponding distributions have 
to be steeper.

We have then repeated all the calculations for the three
different $ASY-EOS$ and we see, as reported in Table I, a nice
increase (modulus) of the isoscaling parameters with the
increasing stiffness of the symmetry energy.

\begin{table}[t]
\begin{center}
\begin{tabular}{|l|c|c|c|} \hline 
       & $asysoft$     & $asystiff$  & $superasystiff$   \\ \hline\hline
   $\alpha$ &   0.69   &     0.95     &   1.05       \\ \hline
   $\beta$  &  -0.67   &    -1.07     &  -1.18        \\ \hline
\end{tabular}
\end{center}
\vskip 0.2cm
\caption{\label{table1} 
The isoscaling parameters $\alpha$ and $\beta$ in neck fragmentation
for three asy-EOS.}
\end{table}

We can summarize, saying that neck fragmentation mechanism, owing to
its particular features, can provide distinct opportunities to
study the density dependence of $asy-EOS$.

\section{CONCLUSIONS}

In the framework of a microscopic transport model, 
based on a stochastic extension of the $BNV$ equation, 
we investigate the ``dynamical'' $IMF$ production 
observed experimentally
in semicentral heavy ion collisions in the Fermi energy domain. 
For the reaction
$^{124}Sn+^{64}Ni$, we identify two main mechanisms
responsible for the IMF emission that cannot be ascribed to the statistical 
decay of PLF or TLF:  
a neck fragmentation
mechanism, a fast process, that
takes place within first $200fm/c$ to $300fm/c$ and the
dynamically induced fission, triggered
by the large deformations acquired during the interaction stage,
that we observe, in some events, at larger times. 

We have studied in detail the features of neck fragmentation, as a source of
$IMF$ emission in the midrapidity region.

This IMF emission appears largely decorrelated  to PLF or TLF statistical
emission. Indeed 
the relative asymptotic velocities $IMF-PL(TL)F$ show deviations from the
Viola systematics ranging from $25\%$ to $70\%$. 
We interpret this as the consequence of an early decoupling of the neck zone,
where the $IMF$'s are formed, from the $PLF$ and $TLF$. Indeed, 
at the separation time,
because of incomplete dissipation of their collective motion,
$PL$ and $TL$ residues have relative velocities,
with respect to the participant region, well 
 above the values associated with  
a pure Coulomb repulsion.


We observe that 
the neck region,  due to expansion effects, as well as to the contemporary 
PLF-TLF centrifugal motion is quenched inside the spinodal instability region.
Hence we identify the spinodal decomposition as the kinetic process
driving fragment formation in the neck region. 
As a consequence, 
we find that the mean field compressibility is governing 
the neck fragmentation dynamics, while the nucleon-nucleon cross section value
shapes its properties.
 
Results regarding the NOF's kinematical properties are 
in qualitative agreement 
with the experimental observations. We can also underline that a proper
confrontation of these events with the  transport model predictions
can impose new constraints on the in medium nucleon-nucleon
cross sections.

We observe interesting effects in  
the isospin dynamics, related to the density and asymmetry gradients
between participant and spectator regions. The neck region becomes more
charge asymmetric due to neutron migration, a process influenced by the 
asy-EOS density dependence. We show that the effect is more pronounced 
for larger slopes
of the symmetry energy around and below saturation density values.
The IMF isospin content keeps track of this early neutron enrichement
process. 

Predictions are in agreement with the experimental
observations, based on isotopic distributions of light particles,
of a neutron richer neck zone. Interesting enough, we observe isoscaling
in the neck fragmentation, inspite
of the short time scales of the IMF production. 
The isoscaling parameters, $\alpha$ and $\beta$, are sensitive to the density
dependence of the asy-EOS, increasing in absolute values
when going from asy-soft to
asy-stiffer EOS.

\subsection*{Acknowledgements}

We warmly thank the $CHIMERA-REVERSE$ Collaboration for the continous
exchange of information on their data analysis. In particular we
a very grateful for the stimulating discussions with Janusz
Wilczynski, Eric Piacsecki, Angelo Pagano and Enrico De Filippo.

\end{document}